\documentclass[10pt, conference, letterpaper]{IEEEtran}
\IEEEoverridecommandlockouts
\usepackage{cite}
\usepackage{amsmath,amssymb,amsfonts}
\usepackage{algorithmic}
\usepackage{graphicx}
\usepackage{subcaption}  
\usepackage{textcomp}
\usepackage{xcolor}
\usepackage[colorlinks,
            linkcolor=blue,
            anchorcolor=blue,
            citecolor=blue]{hyperref}       
\def\BibTeX{{\rm B\kern-.05em{\sc i\kern-.025em b}\kern-.08em
    T\kern-.1667em\lower.7ex\hbox{E}\kern-.125emX}}

\DeclareRobustCommand*{\IEEEauthorrefmark}[1]{%
    \raisebox{0pt}[0pt][0pt]{\textsuperscript{\footnotesize\ensuremath{#1}}}}

\begin{document}

\title{Codebook-enabled Generative End-to-end Semantic Communication Powered by Transformer\\
}


\author{\IEEEauthorblockN{Peigen Ye\IEEEauthorrefmark{1, 2}, Yaping Sun\IEEEauthorrefmark{2, 3}, 
Shumin Yao\IEEEauthorrefmark{2}, 
Hao Chen\IEEEauthorrefmark{2}, 
Xiaodong Xu\IEEEauthorrefmark{4, 2}, 
Shuguang Cui\IEEEauthorrefmark{3, 2}, \emph{Fellow, IEEE}
}
\IEEEauthorblockA{\IEEEauthorrefmark{1}The Shenzhen Campus of Sun Yat-sen University, Sun Yat-sen University, Shenzhen, China}
\IEEEauthorblockA{\IEEEauthorrefmark{2}Peng Cheng Laboratory, Shenzhen, China}


\IEEEauthorblockA{\IEEEauthorrefmark{3}FNii and SSE, The Chinese University of Hong Kong, Shenzhen, China}


\IEEEauthorblockA{\IEEEauthorrefmark{4}Beijing University of Posts and Telecommunications, Beijing, China}
}


\maketitle

\begin{abstract}
Codebook-based generative semantic communication attracts increasing attention, since only indices are required to be transmitted when the codebook is shared between transmitter and receiver. 
However, due to the fact that the semantic relations among code vectors are not necessarily related to the distance of the corresponding code indices, the performance of the codebook-enabled semantic communication system is susceptible to the channel noise. Thus, how to improve the system robustness against the noise requires careful design. 
This paper proposes a robust codebook-assisted image semantic communication system, where semantic codec and codebook are first jointly constructed, and then vector-to-index transformer is designed guided by the codebook to eliminate the effects of channel noise, and achieve image generation. 
Thanks to the assistance of the high-quality codebook to the Transformer, the generated images at the receiver outperform those of the compared methods in terms of visual perception. In the end, numerical results and generated images demonstrate the advantages of the generative semantic communication method over JPEG+LDPC and traditional joint source channel coding (JSCC) methods.

\end{abstract}


\section{Introduction}

Semantic communications have recently been recognized as a promising technology for beyond 5G (B5G) and 6G wireless networks, whereby transmitters are designed to efficiently convey semantic information to receivers, rather than reliably transmit syntactic information as in conventional wireless communication systems. Via the end-to-end (E2E) semantic information transmission design, the semantic communications are able to efficiently compress messages while preserving the essential meaning by filtering out the irrelevant information, thus significantly enhancing the communication efficiency. Semantic communications are envisioned to have abundant applications such as network intelligence  and industrial automation. 

Artificial intelligence technology significantly contributes to the progress of communication technology \cite{9651548} \cite{10040976} \cite{10175391}. Semantic communication relies primarily on artificial intelligence techniques to 
construct a series of task-specific or general-purpose semantic knowledge bases. 
Semantic knowledge base (SKB), as a key enabler of semantic communication, facilitates semantic encoding/decoding via providing semantic knowledge vectors, and refining a compact search space \cite{ren2023knowledge}.
For instance, Sun et al. \cite{sun2023semantic} proposed a novel SKB-enabled multi-level feature transmission framework that significantly improves the performance of remote zero-shot transmission. Li et al. \cite{li2022domain} introduced a medical semantic communication system, which leverages a domain knowledge with retinal fundus segmentation labels to enhance image reconstruction accuracy at the receiver.

Regarding the construction of SKBs, there is a current trend towards 
discrete quantized codebooks. For example, Hu et al. \cite{hu2023robust} proposed a semantic communication method with masked vector quantized-variational autoencoder (VQ-VAE) enabled codebook. This method utilizes a discrete codebook to share encoded feature representations between the transmitter and the receiver. Park et al. \cite{9952500} proposed a novel DeepSC framework with federated codebook to serve multi-user scenarios \cite{10146016}. The codebook of discrete coding serves as apriori knowledge, endowing the encoder with a more stable mapping target range during the encoding process. Simultaneously, it reduces the bias in information comprehension between the encoder and decoder. This not only enhances the stability of feature encoding by the encoder, but also significantly improves the transmission efficiency.

However, when indices are transmitted over wireless channel, a minor error of decoded indices at the receiver could potentially lead to catastrophic consequences for the reconstruction of the original information. This is because each index generally corresponds to a substantial amount of semantic information, and the semantic relations among code vectors are not necessarily related to the distance of the corresponding code indices.
As an alternative approach, transmitting the feature map at the transmitter and subsequently performing codebook-based feature matching at the receiver could be a viable option. However, the transmission of vector features is also susceptible to the noise in the physical communication channel. Although the impact of feature contamination by noise may be relatively smaller compared to the tampering of indices, it remains a factor that requires careful consideration.

In recent years, the development of Transformer \cite{vaswani2017attention} has been widely recognized, particularly with the emergence of large models such as ChatGPT, which has propelled artificial intelligence to new heights. The self-attention mechanism employed by Transformer allows each position in the input sequence to dynamically attend to all other positions, enabling the model to directly gather information from the entire input sequence. This emphasis on global information over local details endows Transformer with outstanding performance in various tasks, especially in situations that require consideration of long-range dependencies. In \cite{zhou2022towards}, Zhou et al. addressed the matching problem between degraded image feature vector and high-quality image feature vector by introducing the Transformer. Inspired by their work \cite{zhou2022towards} \cite{esser2021taming}, this paper incorporates the Transformer to tackle the challenge of recovering image feature map, as transmitted information, after being contaminated by noise. The goal is to restore these features to the state of indices suitable for image reconstruction.

Main contributions of this paper lie in the following aspects. 

\begin{itemize}
\item We build a codebook-assisted generative semantic encoding and decoding architecture. And in the receiver, a transformer with nine self-attention blocks is used to realize the mapping of feature map with errors to high-quality codebook indices based on the global information of the feature map and the assistance of codebook, thereby eliminating noise contamination.
\item The proposed method adopts a two-stage training mechanism. In Stage 1, the encoder, decoder, and codebook are obtained through E2E training. In Stage 2, the vector-to-index Transformer is fully trained by fixing the upstream and downstream parameters.
\item Simulation results and generated images demonstrate the advantages of generative semantic communication methods over JPEG+LDPC and traditional joint source channel coding (JSCC) methods. Under low signal-to-noise ratio (SNR), the proposed method can still maintain good performance on the Learned Perceptual Image Patch Similarity (LPIPS) metric and perform high-quality image transmission.
\end{itemize}

\section{System Model}
In this section, we present the framework of codebook-enabled generative E2E semantic communication system. As illustrated in Fig. \ref{fig1:system}, the input image is first encoded into a feature map by the semantic encoder at the transmitter, and then the feature map is transmitted through the wireless channel. Subsequently, the received feature map is first mapped into the corresponding indices by the vector-to-index Transformer (V2IT), and then decoded by the semantic decoder to generate a high-quality image with similar visual perception. The codebook $\mathcal{C}$ consists of a set of image features ${c_k}$, expressed as $\mathcal{C} = \left\{ {{c_k} \in {\mathbb{R}^d}} \right\}_{k = 0}^L$.


\subsection{Semantic Encoder at Transmitter}

\begin{figure}[t]  
  \centering
  \includegraphics[width=3.4in]{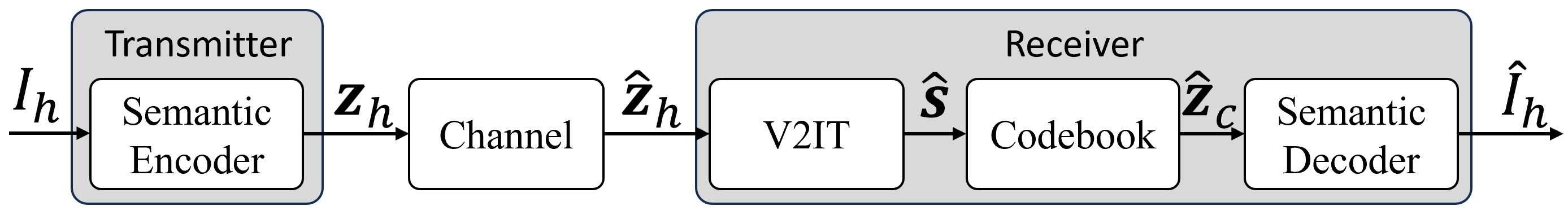}
  \caption{The framework of the proposed semantic communication system.}
  \label{fig1:system}
\end{figure}

As shown in Fig. \ref{fig1:system}, the transmitter is endowed with an encoder with a Convolutional Neural Network (CNN) structure, designed to extract high-dimensional spatial features from the input image. Initially, the input image ${I_h} \in {\mathbb{R} ^{H \times W \times 3}}$ is fed into the network and normalized. Subsequently, within the encoder, after undergoing operations such as convolution, the image is encoded into a feature vector $ {\mathbf{z}_h} \in {\mathbb{R} ^{m \times n \times q}}$. This process can be represented as follows:

\begin{equation}
{\mathbf{z}_h} = {\mathbf{E}_\mathbf{\theta} }\left( {{I_h}} \right){,}
\end{equation}
where $\mathbf{E}_\mathbf{\theta}\left ( \cdot  \right )$ represents the entire semantic encoding function with respect to parameters $\mathbf{\theta}$.

Due to the first stage of training, where the encoder contributes to the generation of the codebook through end-to-end training, the encoded feature map ${\mathbf{z}_h}$ is highly similar to the vector in the codebook. ${\mathbf{z}_h}$ is composed of $m \times n$ $q$-dimensional feature vector ${z_h^{\left( {i,j} \right)}}$. Its overall size is $N = m \times n \times q$, which is closely related to the amount of transmitted information. In fact, the semantic encoder can be considered a generalized SKB, as it learns the knowledge of encoding the image to vectors that approximate the vectors in the codebook. The performance of the encoder also affects the recovery of the feature map in the subsequent processes.

\begin{figure*}[t]  
  \centering
  \includegraphics[width=6.1in]{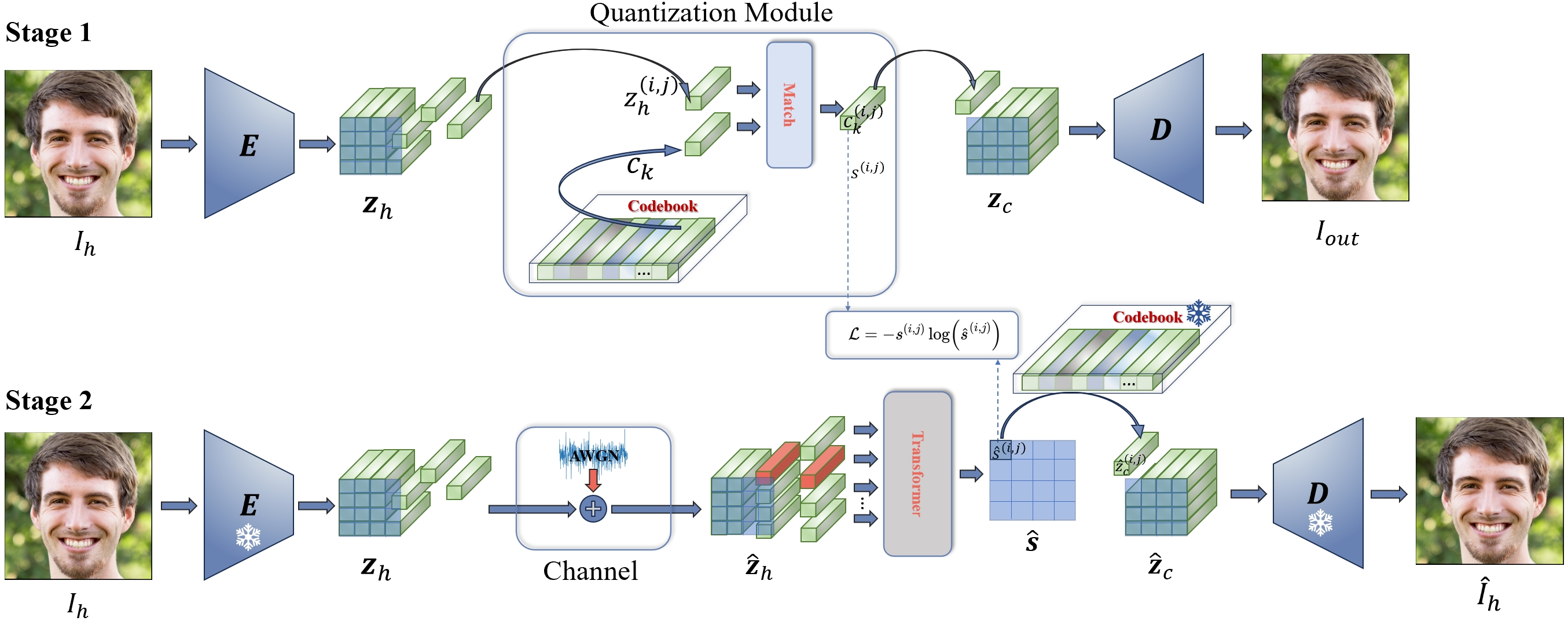}
  \caption{The network architecture of the proposed semantic communication system, a two-stage trained neural network model. In Stage 1, a high-quality codebook, enriched with semantic details from trained images, is constructed through learning. Simultaneously, an encoder and a decoder are obtained. In Stage 2, the Transformer, as the vector-to-index Transformer, is trained by fixing its upstream and downstream parameters. The corrected feature map are utilized for image reconstruction.}
  \label{fig:two-stage network}
\end{figure*}

\subsection{Physical Communication Channel}

Assuming a single communication link for image semantic communication transmission, to simulate the process of physical information transmission, we have employed the Additive White Gaussian Noise (AWGN) channel model widely used in wireless communication as the physical communication channel.
${\mathbf{z}_h}$ undergoes transmission through the physical channel to reach the receiver.
The signal received at the receiver is denoted as ${\hat{\mathbf{z}}_h}$, and this process can be represented as follows:

\begin{equation}
{\hat{\mathbf{z}}_h} = {\mathbf{z}_h} + \mathbf{n}{,}
\end{equation}
where the AWGN $\mathbf{n}$ has i.i.d elements with zero mean and variance ${\sigma ^2}$.


\subsection{Vector-to-index Transformer, Codebook, and Semantic Decoder at Receiver}
\label{sec:2rd_section_Receiver}

The corrupted output ${\hat{\mathbf{z}}_h}$ from the wireless communication physical channel is acquired by the receiver. Initially, in the receiver, vector-to-index Transformer predicts and corrects the feature map based on its global information, resulting in the semantic feature indices $\hat{\mathbf{s}}$. This process can be expressed as follows:

\begin{equation}
\hat{\mathbf{s}} = {\mathbf{T}_\mathbf{\tau} }\left( {\hat{\mathbf{z}}_h} \right){,}
\end{equation}
where ${\mathbf{T}_\mathbf{\tau} }\left( \cdot \right)$ indicates the function of vector-to-index Transformer with respect to parameters $\mathbf{\tau}$.

Next, the system can retrieve the corresponding feature map from the codebook based on the feature indices $\hat{\mathbf{s}} = {\left\{ {{{\hat s}^{\left( {i,j} \right)}}} \right\}^{m \cdot n}}$. This retrieval process reconstructs the complete feature map $ {\hat{\mathbf{z}}_c} \in {\mathbb{R} ^{m \times n \times q}}$ . This process can be expressed as:

\begin{equation}
\hat{z}_c^{\left( {i,j} \right)} = \mathbf{S}\left( {{{\hat s}^{\left( {i,j} \right)}}} \right){,}
\end{equation}
where $\mathbf{S}\left( \cdot \right)$ denotes the method of retrieval (Look-up Table), ${\hat s^{\left( {i,j} \right)}} \in \left\{ {0, \cdots, L - 1} \right\}$, and $L$ represents the size of the codebook.

Subsequently, the feature map ${\hat{\mathbf{s}}_c}$ is fed into the decoder for image reconstruction, resulting in the output ${\hat{I}_h}$. This process can be expressed as:

\begin{equation}
{\hat I_h} = {\mathbf{D}_\mathbf{\xi} }\left( {\hat{\mathbf{z}}_c} \right){,}
\end{equation}
where ${\mathbf{D}_\mathbf{\xi} }\left( \cdot \right)$ represents the decoder function with respect to parameters $\mathbf{\xi}$, which is an image generation module.


\section{The Proposed Method}
\label{sec:third_section}
The image feature vector is contaminated by noise when it passes through the channel, causing confusion in image reconstruction. Therefore, the core objective of our work is to recover and correct the contaminated feature vector, mapping it to a vector index from the learned codebook that is more semantically related to the original feature vector.

This section introduces the specific training method of the proposed semantic communication system. The training process of the proposed method consists of two stages. In Stage 1, through end-to-end training on image dataset, it generates a codebook and an encoder-decoder pair. In Stage 2, it addresses the challenge of how the image feature map, contaminated by channel noise, can be corrected as accurately as possible by its global information and the codebook.


\subsection{Stage 1: Joint Construction of Semantic Codec and Codebook}\label{Stage1}

To obtain the codebook and the encoder-decoder pair, a vector-quantized autoencoder network \cite{esser2021taming} is constructed. As illustrated in Fig. \ref{fig:two-stage network}, in Stage 1, a high-quality input image ${I_h} \in {\mathbb{R} ^{H \times W \times 3}}$ is fed into the encoder $\mathbf{E}$, resulting in the encoded feature map $ {\mathbf{z}_h} \in {\mathbb{R} ^{m \times n \times q}}$. Subsequently, it is passed through the quantization module.
In the quantization module, each vector $z_h^{\left( {i,j} \right)}$ in ${\mathbf{z}_h}$ is matched to the quantized vector ${c_k}$ from the codebook $\mathcal{C}$ using nearest-neighbor matching. Subsequently, the quantization module places the matched vector $c_k^{\left( {i,j} \right)}$ at the corresponding position ${\left( {i,j} \right)}$, combines them to obtain the map ${\mathbf{z}_c}$ for reconstruction, and simultaneously outputs the index 
$s^{\left( {i,j} \right)}$ ($\mathbf{s} = {\left\{ {{s^{\left( {i,j} \right)}}} \right\}^{m \cdot n}}$) corresponding to $c_k^{\left( {i,j} \right)}$. Finally, the ${\mathbf{z}_c}$ is input into the decoder (image generation module) for image reconstruction, resulting in the reconstructed image ${I_{out}}$.

Training Objectives. For the E2E network in Stage 1, based on the network's input and output, we can establish a result-based loss function $\mathcal{L}_{result}$:

\begin{equation}
\mathcal{L}_{result} = {\left\| {{I_h} - {I_{out}}} \right\|_1} + \left\| {\mathbf{\phi} \left( {{I_h}} \right) - \mathbf{\phi} \left( {{I_{out}}} \right)} \right\|_2^2{,}
\end{equation}
where ${\left\| {{I_h} - {I_{out}}} \right\|_1}$ represents the L1 loss, which measures the mean absolute difference between the output $I_{out}$ and the ground truth $I_h$. $\left\| {\mathbf{\phi} \left( {{I_h}} \right) - \mathbf{\phi} \left( {{I_{out}}} \right)} \right\|_2^2$ denotes the perceptual loss, reflecting the differences between the output and the ground truth in a high-dimensional space, where $\mathbf{\phi} \left( \cdot \right)$ represents the deep feature extractor, utilizing the VGG19 \cite{simonyan2014very}.

Due to the insufficient constraints of the above loss functions on the quantization module, the loss function $\mathcal{L}_{in}$ reflecting the intermediate layers of the network is adopted:

\begin{equation}
\mathcal{L}_{in} = \left\| {\mathbf{sg}\left( {{\mathbf{z}_h}} \right) - {\mathbf{z}_c}} \right\|_2^2 + \beta \left\| {{\mathbf{z}_h} - \mathbf{sg}\left( {{\mathbf{z}_c}} \right)} \right\|_2^2{,}
\end{equation}
where ${\mathbf{sg}}\left( \cdot \right)$ represents the stop-gradient operator, and ${\beta}$ is the weight parameter for the update rates of the encoder and the codebook.

As an adversarial model is employed as the generator module, the overall loss function needs to include an adversarial loss function $\mathcal{L}_{GAN}$:

\begin{equation}
\mathcal{L}_{Stage1} = \mathcal{L}_{result}+\mathcal{L}_{in}+\mathcal{L}_{GAN}{,}
\end{equation}


\subsection{Stage 2: Vector-to-index Transformer}
In Stage 2, our goal is to obtain the vector-to-index Transformer, which can utilize the codebook and the global information of the contaminated feature map to correct feature map, to obtain vector indices from the learned codebook that are most semantically related to the original image features. The network architecture of Stage 2 was introduced in the Section \ref{sec:2rd_section_Receiver}. 

Training Objectives. The input of Transformer is the contaminated image feature map, and the output is the indices from the codebook. Based on the training results of Stage 1, we can formulate the loss function for Stage 2 as follows:

\begin{equation}
\mathcal{L}_{Stage2} = \lambda \sum\limits_{i = 0}^{mn - 1} {\left( { - {{s^{\left( {i,j} \right)}}}\log \left( {{{{\hat s}^{\left( {i,j} \right)}}}} \right)} \right)}  + \left\| {{{\hat {\mathbf{z}}}_h} - \mathbf{sg}\left( {{\mathbf{z}_c}} \right)} \right\|_2^2{,}
\end{equation}
where $s^{\left( {i,j} \right)}$ and ${\mathbf{z}_c}$ are both derived from Stage 1. They are considered as the ground truths for the corresponding variables in the Stage 2 network. $\lambda$  represents a weight parameter. During the training process of Stage 2, the parameters of the codebook, encoder and decoder are kept unchanged, and only the vector-to-index Transformer is trained. Towards the end of the training, all parameters can be unfrozen to fine-tune the network.

\section{Experiment and Evaluation}

\subsection{Datasets and Settings}

\textbf{The training dataset:} it consists of 68,659 high-quality images sourced from the FFHQ dataset \cite{karras2019style}. The native resolution of images in the FFHQ dataset is $1024 \times 1024$. During dataset preparation, the images are preprocessed, and their sizes are resized to $512 \times 512$.

\textbf{Two test datasets:} the FFHQ-test comprises 1,094 high-quality images from the FFHQ dataset (non-overlapping with the training dataset), and the CelebA-test includes 304 high-quality images from the CelebA dataset \cite{karras2017progressive}. The sizes of all images in both datasets are adjusted to $512 \times 512$.

\textbf{Settings:} the size of the Codebook is set to 1024 (i.e., $L=1024$). The length of vector in the codebook is set to 256 (i.e., $q=256$). For all training stages, we employ the Adam 
optimizer with a batch size of 4. The learning rates for Stage I and Stage II are set to $7 \times {10^{ - 5}}$ and $1 \times {10^{ - 4}}$, respectively. The training iterations for Stage 1 and Stage 2 are set to $415 \times 10^3$ and $200 \times 10^3$, respectively. The proposed method is implemented based on the PyTorch framework and trained using one GeForce RTX 3090 GPU.

\subsection{Simulation and Evaluation}

\begin{figure*}[ht]
  \centering

  \begin{subfigure}{0.3\textwidth}
    \includegraphics[width=\linewidth]{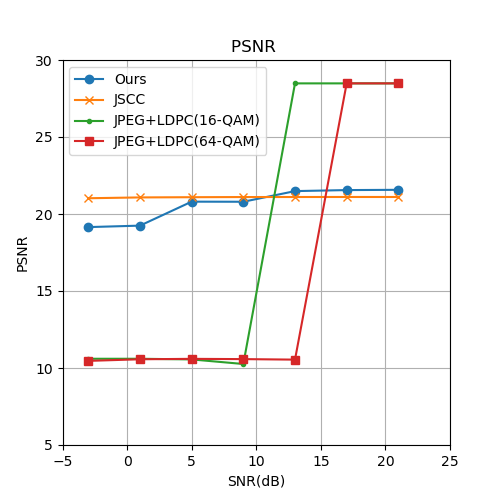}
    \caption{SNR-PSNR in FFHQ-test dataset}
    \label{fig:subfig1}
  \end{subfigure}
  \hfill
  \begin{subfigure}{0.3\textwidth}
    \includegraphics[width=\linewidth]{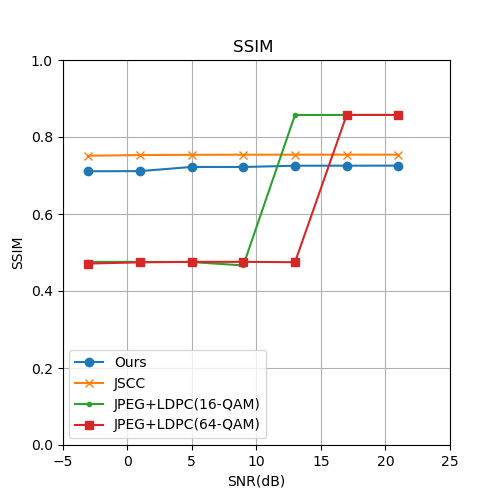}
    \caption{SNR-SSIM in FFHQ-test dataset}
    \label{fig:subfig2}
  \end{subfigure}
  \hfill
  \begin{subfigure}{0.3\textwidth}
    \includegraphics[width=\linewidth]{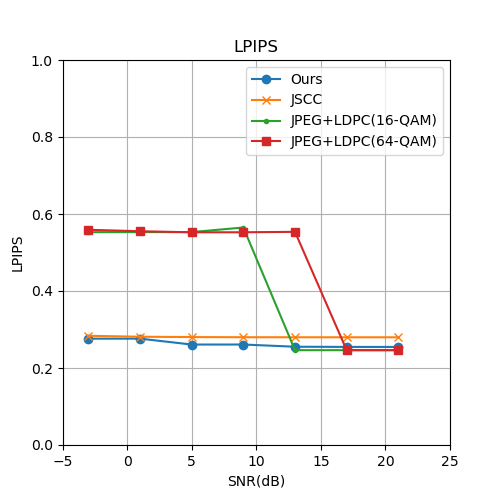}
    \caption{SNR-LPIPS in FFHQ-test dataset}
    \label{fig:subfig3}
  \end{subfigure}

  \begin{subfigure}{0.3\textwidth}
    \includegraphics[width=\linewidth]{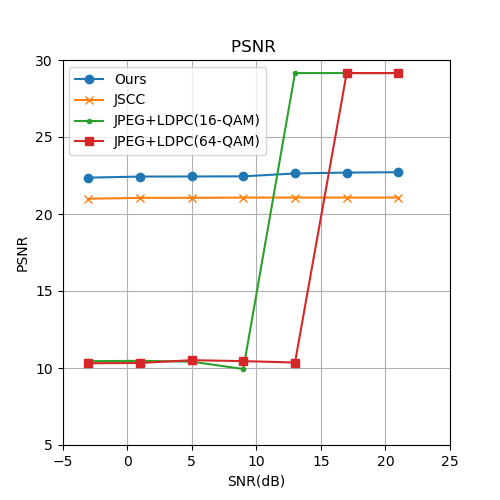}
    \caption{SNR-PSNR in CelebA-test dataset}
    \label{fig:subfig4}
  \end{subfigure}
  \hfill
  \begin{subfigure}{0.3\textwidth}
    \includegraphics[width=\linewidth]{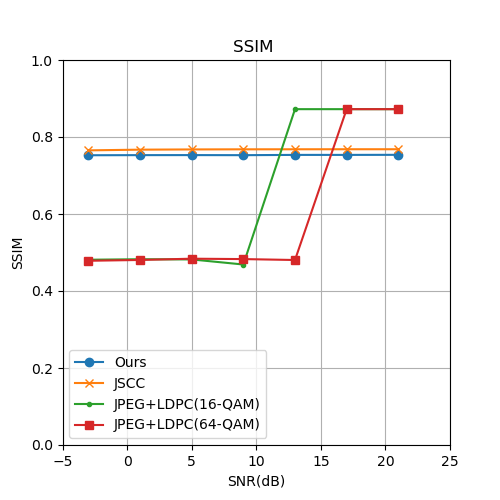}
    \caption{SNR-SSIM in CelebA-test dataset}
    \label{fig:subfig5}
  \end{subfigure}
  \hfill
  \begin{subfigure}{0.3\textwidth}
    \includegraphics[width=\linewidth]{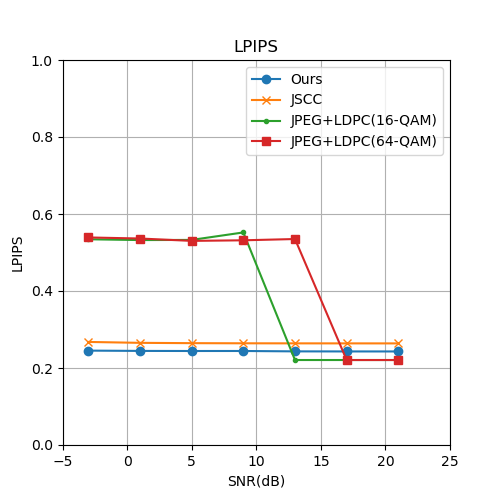}
    \caption{SNR-LPIPS in CelebA-test dataset}
    \label{fig:subfig6}
  \end{subfigure}

  \caption{The performance metrics, including PSNR, SSIM, and LPIPS, for three methods across the SNR range of -3 to 21. The evaluation is conducted on the FFHQ-test dataset and CelebA-test dataset.}
  \label{fig:3}
\end{figure*}

In this section, the proposed method will be validated and compared with two other excellent methods: JPEG+LDPC and JSCC. 

\textbf{JPEG+LDPC.} JPEG, established by the Joint Photographic Experts Group, is a widely used traditional image coding standard. LDPC, a powerful error correction coding method, is commonly employed in channel coding for digital communication systems. 
Communication systems combining JPEG image coding and LDPC channel coding are generally considered to have high reliability in image transmission, especially in communication environments affected by noise. In the following experiments, the bit rate is set as $1/2$.

\textbf{JSCC.} JSCC \cite{bourtsoulatze2019deep} is a classical joint source-channel coding technique in the field of semantic communication, which parameterizes the encoder and decoder functions by two CNNs and trains them jointly.

About the experimental evaluation, we not only calculate traditional image quality assessment metrics such as PSNR (Peak Signal-to-Noise Ratio) and SSIM (Structural Similarity), but also include the LPIPS (Learned Perceptual Image Patch Similarity). Due to employing a pre-trained deep feature extractor, LPIPS is more agreeable with human perceptual judgments \cite{li2022domain}.

\begin{figure*}[t]  
  \centering
  \includegraphics[width=6.6in]{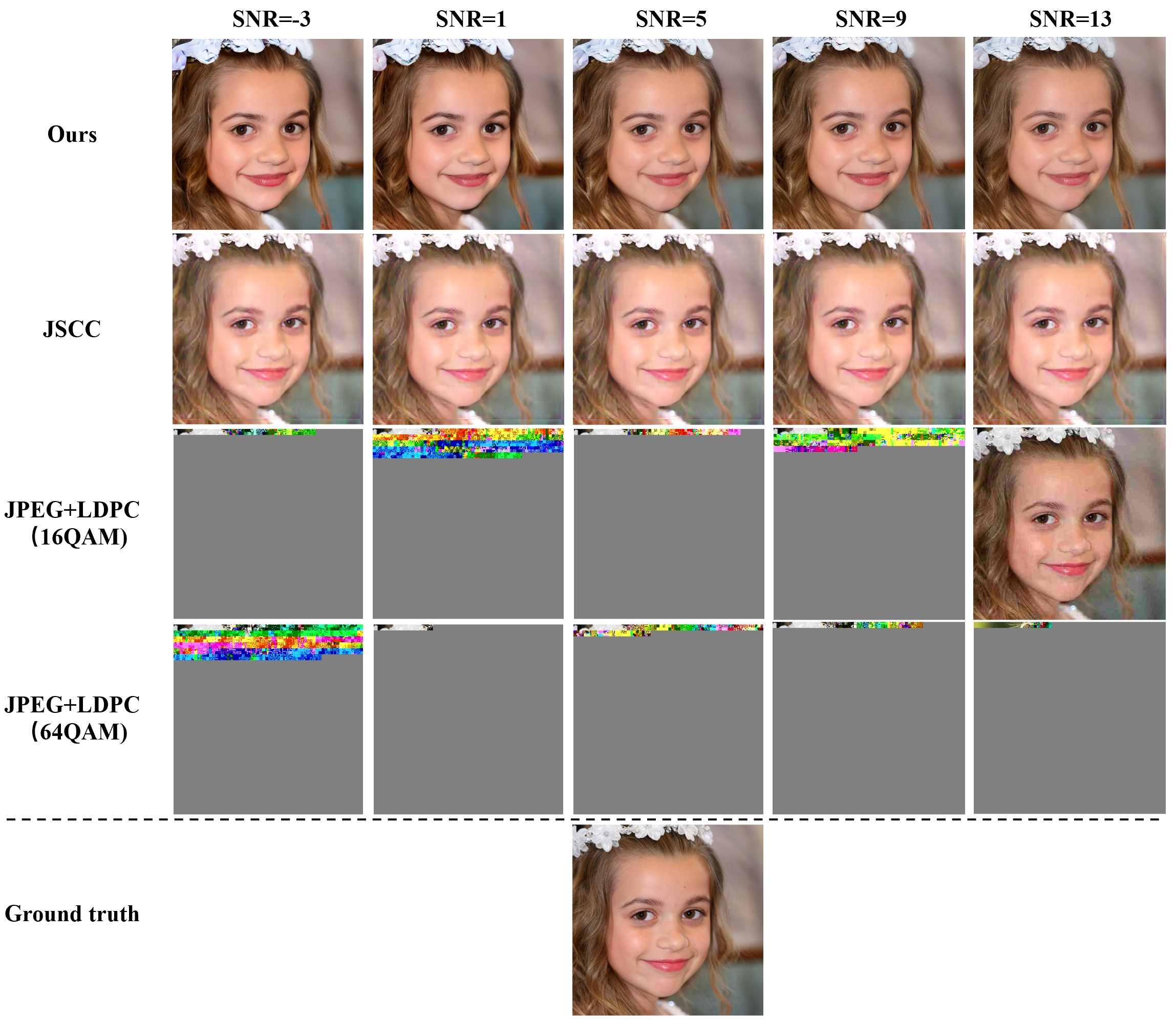}
  \caption{The output results of image transmission for three methods under different SNR values (-3, 1, 5, 9, 13). The images represent the output of image transmission for each method. The gray images in the JPEG+LDPC method column represent cases where information loss is severe, making it challenging to decode normal images.}
  \label{fig:Pic_show}
\end{figure*}

Fig. \ref{fig:3} illustrates the variations in the PSNR, SSIM, and LPIPS metrics for the three image transmission methods under different SNR on FFHQ-test and CelebA-test datasets. Fig. \ref{fig:Pic_show} depicts the actual input and output images corresponding to the respective situations.

From each subplot in Fig. \ref{fig:3}, it is evident that the JPEG+LDPC method, representing traditional communication approaches, exhibits a drastic performance decline in image transmission under low SNR environments. This decline is observed across all metrics (where LPIPS ranges from 0 to 1, with smaller values indicating higher perceptual similarity between two images). In Fig. \ref{fig:Pic_show}, it is more visually apparent that JPEG+LDPC suffers severe information loss in low SNR scenarios, to the extent that image decoding becomes practically impossible. The gray images in Fig. \ref{fig:Pic_show} are contingent on our a specific operation. In simulation experiments, we specifically provided lossless data associated with the JPEG decoding protocol to JPEG+LDPC at the receiver. Without this operation, even the gray images would not be obtainable.

From Fig. \ref{fig:subfig1}, Fig. \ref{fig:subfig2}, and Fig.  \ref{fig:subfig5}, it can be observed that when SNR is high (e.g., SNR $>$ 12dB), our proposed method exhibits slightly lower PSNR and SSIM compared to the JSCC method. However, our method consistently outperforms the comparison method in terms of LPIPS. Moreover, as depicted output results in Fig. \ref{fig:Pic_show}, the obtained images using our method visually outperform those of the comparison methods significantly. This is because PSNR and SSIM are metrics based on the pixel level, evaluating images in a low-dimensional space. Such comparisons are not very suitable for semantic communication tasks, especially generative semantic communication. 
This is because the characteristic of generative communication network, i.e., the unavoidable randomness within a certain range, makes it naturally less advantageous for pixel-wise or low-dimensional structural comparisons like PSNR and SSIM. However, for the majority of semantic communication tasks, the focus is on transmitting crucial semantic information required for the task, rather than pursuing absolute accuracy for each pixel. Hence, metrics resembling per-pixel comparisons may not hold significant importance in semantic communication.


\section{Conclusion}
This paper proposes a robust codebook-assisted image semantic communication method, via transmitting the feature map directly to the receiver, and leveraging Transformer to predict and correct feature map with errors based on the global information of the feature map and the assistance of codebook, achieving excellent source image reconstruction.
The use of a codebook significantly enhances the performance of vector-to-index Transformer, thereby improving communication stability and efficiency. Furthermore, images generated based on high-quality map from the learned codebook surpass those of comparative methods in terms of visual aesthetics. Additionally, we conduct a thorough analysis of the characteristics of generative semantic communication in image transmission, elucidating the superiority of generative semantic communication methods. This work further substantiates the promising prospects and value of generative network structures in the field of semantic communication.

\bibliographystyle{unsrt}
\bibliography{IEEEexample}



\end{document}